\documentclass[aps,prb,10pt,twocolumn,superscriptaddress,nofootinbib,preprintnumbers,showpacs]{revtex4-2}

\usepackage{amsmath,amsfonts, amssymb, amsthm, dsfont}
\usepackage[T1]{fontenc}
\usepackage{yfonts}
\usepackage{bm}
\usepackage{bbm}
\usepackage{mathrsfs}
\usepackage{graphicx}
\usepackage{verbatim}
\usepackage[bookmarks=true,colorlinks=true,linkcolor=blue, urlcolor=blue, citecolor=blue]{hyperref}
\hypersetup{linktocpage}
\usepackage{wasysym}
\usepackage[dvipsnames]{xcolor}
\usepackage{bbold}
\usepackage{braket}
\usepackage{diagbox} 
\usepackage{environ} 
\usepackage{relsize} 
\usepackage[caption=false]{subfig} 
\usepackage{tikz}
\usepackage{ulem}

\usepackage[dvipsnames]{xcolor}

\newcommand{\mysec}[1]{\textit{{\textcolor{black}{#1}}}}

\begin{document}
\title{\mbox{Scaling of the disorder operator at (3+1)D O(3) quantum criticality}} 

\author{Xuyang Liang}
\affiliation{Guangdong Provincial Key Laboratory of Magnetoelectric Physics and Devices, State Key Laboratory of Optoelectronic Materials and Technologies, Institute of Neutron Science and Technology, School of Physics, Sun Yat-Sen University, Guangzhou, 510275, China}

\author{Xiao-Chuan Wu}
\email{xw0618@princeton.edu}
\affiliation{Department of Physics, Princeton University, Princeton, NJ 08544, USA}
\affiliation{Leinweber Institute for Theoretical Physics, University of Chicago, Chicago, IL 60637, USA}

\author{Zenan Liu}
\affiliation{Department of Physics, South China University of Technology, Guangzhou 510640, China}
\affiliation{Department of Physics, School of Science and Research Center for Industries of the Future, Westlake University, Hangzhou 310030,  China}
\affiliation{Institute of Natural Sciences, Westlake Institute for Advanced Study, Hangzhou 310024, China}

\author{Zhe Wang}
\affiliation{Department of Physics, School of Science and Research Center for Industries of the Future, Westlake University, Hangzhou 310030,  China}
\affiliation{Institute of Natural Sciences, Westlake Institute for Advanced Study, Hangzhou 310024, China}

\author{Zheng Yan}
\email{zhengyan@westlake.edu.cn}
\affiliation{Department of Physics, School of Science and Research Center for Industries of the Future, Westlake University, Hangzhou 310030, China}
\affiliation{Institute of Natural Sciences, Westlake Institute for Advanced Study, Hangzhou 310024, China}

\author{Dao-Xin Yao}
\email{yaodaox@mail.sysu.edu.cn}
\affiliation{Guangdong Provincial Key Laboratory of Magnetoelectric Physics and Devices, State Key Laboratory of Optoelectronic Materials and Technologies, Institute of Neutron Science and Technology, School of Physics, Sun Yat-Sen University, Guangzhou, 510275, China}

\begin{abstract}
The disorder operator, as an easily measured nonlocal observable, displays great potential in detecting intrinsic information of field theories. It has been systematically studied in one- and two-dimensional (1D and 2D) quantum systems, while the knowledge of 3D is still limited. The disorder operator associated with U(1) global symmetry exhibits rich geometric dependence on the shape of the spatial region at a quantum critical point, meanwhile, (3+1)D is the upper critical dimension for $O(N)$ criticality, both of which pose a challenge for exploring the disorder operator in high dimensions. 
In this Letter, we investigate the scaling behaviors of disorder operators in (3+1)D $O(3)$ models through large-scale quantum Monte Carlo simulation combined with theoretical analysis. Although the upper critical dimension introduces logarithmic corrections in correlations, we analytically prove and numerically demonstrate that the corrections do not modify the universal trihedral-corner contribution of the disorder operator.
The universal contributions, such as the current central charge, have been revealed in our calculation, which establishes a concrete link between lattice simulations and continuum field theory. This work opens promising directions for the experimental and numerical exploration of universal properties at quantum critical points in (3+1)D models.

\end{abstract}


\date{\today}

\maketitle

\mysec{Introduction.}
In recent years, nonlocal observables have substantially deepened our understanding of quantum phases and phase transitions, both conceptually and quantitatively. A prominent example is the disorder operator, defined as the expectation value of a symmetry transformation applied to a finite region of a many-body system~\cite{DO_old_1,DO_old_2,DO_fradkin}. On the conceptual side, it provides a unifying framework for a wide class of quantum phases through the lens of generalized symmetries~\cite{Nussinov2006,Nussinov2009,Gaiotto_2015,gene_sym_review1,gene_sym_review2}. On the quantitative side, extensive progress has been made in (2+1)-dimensional [(2+1)D] systems, including studies of spontaneously broken symmetry phases, Landau phase transitions, symmetry-enriched topological orders, symmetric mass generation transitions, Landau Fermi liquids, and quantum critical metals~\cite{zhao2021higher,zhao2021higher, xiao2021universal,estienne2022cornering,Dirac_log,wang2022scaling,liu2022measuring,zhao2022scaling,Fermion2023Liu,Liu2024Disorder,XCWu2020,BBChen2022,jiangFermion2022,Cai_FS_2024,Wu_FS_2024}.

Particularly, in the (2+1)D conformal field theory (CFT), significant progress has been made in relating the scaling of disorder operators to universal CFT data. The dependence on region geometry, such as corner contributions~\cite{Dirac_log,wang2021scaling,xiao2021universal,estienne2022cornering,wang2022scaling,liu2022measuring,zhao2022scaling,Fermion2023Liu,Liu2024Disorder}, closely parallels that of entanglement entropy (EE)~\cite{JRZhao2022,zhao2022scaling,song2024quantum,song2023extracting,deng2023improved,deng2024diagnosing,wang2024ee,wang2024probing,ding2024tracking,jiang2024high} and encodes quantities including the current central charge. These results establish the disorder operator as a complementary and powerful probe of conformal quantum criticality.

Either in theoretical and numerical calculations, or in experimental measurements, the disorder operator is more accessible than the EE~\cite{XCWu2020,BBChen2022,wang2025singularity,sarma2025probing}, although its knowledge is still limited, such as in high dimensions. Therefore, further exploring its connection with field theory is of great significance for promoting the development of related fields.
Meanwhile, in the (3+1)D CFT, geometrical dependence has also played a central role in the study of EE. The greater variety of geometric singularities in higher dimensions leads to a much richer structure of universal terms than in (2+1)D systems.

For smooth entangling surfaces, the universal logarithmic term of EE is determined by the Weyl anomalies~\cite{Solodukhin_2008,CHM_2011}. When the surface develops singular features, additional universal contributions arise, including a double-logarithmic ($\log^2$) term from conical corners~\cite{Myers_Singh_2012,klebanov2012shape,Faulkner2016} and a logarithmic term from trihedral corners~\cite{4D_corner_1,4D_corner_2,4D_corner_3,4D_corner_4,4D_corner_5,EE_singular_2019}. These developments of EE naturally motivate us to investigate whether the disorder operator exhibits analogous geometric dependence in higher dimensions. Moreover, as an equal-time observable, it can be computed at much lower cost than EE, allowing simulations of significantly larger (3+1)D systems and thereby providing more effective control over finite-size effects. Similarly, the 3D fermionic Hubbard model has been successfully constructed using ultracold atoms in optical lattices~\cite{shao2024antiferromagnetic,hart2015observation}, the disorder operator, which can be accessed from snapshots of configurations with polynomial computational cost, serves as a powerful tool to explore the (3+1)D quantum critical phenomena and their associated CFTs while EE needs exponential resources in measurement, e.g., tomography~\cite{steffen2006measurement,kokail2021entanglement}.

In this Letter, we focus on cubic geometry, which is directly relevant to the microscopic lattice models of interest. We begin with a discussion of analytical results for the disorder operator in (3+1)D quantum critical systems, with particular emphasis on the universal logarithmic contribution arising from tetrahedral corners. 
To benchmark these theoretical predictions, we study conventional $\textrm{O}(3)$ symmetry-breaking transitions in the (3+1)D $S=1/2$ Heisenberg double-cubic (DC) and columnar-dimerized (CD) models using large-scale quantum Monte Carlo (QMC) simulations. 
The measured scaling behaviors of the disorder operators across the various phases are in agreement with the corresponding analytical predictions.
We also extract the current central charge at $(3+1)$D quantum critical point, originating from the corner contribution, reflecting the universal CFT information. Our work provides a promising route for exploring $(3+1)$D quantum criticality via the nonlocal operators.

\mysec{Disorder operator.} We begin with general considerations of (3+1)D quantum systems with a U(1) global symmetry. The disorder operator, associated with a spatial subregion $M$, is defined as 
\begin{flalign}
X_{M}(\theta)=\prod_{j\in M}e^{\mathtt{i}\theta n_{j}},
\end{flalign}
where $n_{j}$ denotes the U(1) charge on site $j$, and $\theta$ is a real-valued parameter. In gapped phases, the disorder operator generally follows an area law $X_{M}(\theta)\sim e^{-a_{1}(\theta)l^{2}}$ ($a_1$>0), where $l$ is the linear size of the subregion $M$. When the U(1) symmetry [or a larger group containing U(1)] is spontaneously broken, the area law receives a logarithmic enhancement $X_{M}(\theta)\sim e^{-a_{2}(\theta)l^{2}\ln(l)}$ ($a_2$>0) due to the presence of gapless Goldstone modes~\cite{BF_gas_1998,BF_gas_2007,BF_QCP_2012,Lake_2018,wang2022scaling,wang2021scaling}.

We are particularly interested in quantum criticality. At a conformal critical point, the disorder operator defined over a region $M$ shaped as an $l\times l\times l$ cube is expected to obey the scaling form
\begin{flalign}
\ln|\langle X(\theta)\rangle|=al^{2}+bl+s\ln l+c,
\label{eq:_DO_fit}
\end{flalign}
where all coefficients depend on $\theta$. This expectation is supported by the small-$\theta$ expansion
\begin{flalign}
\langle X_{M}(\theta)\rangle=1-\frac{\theta^{2}}{2}\mathcal{C}_{M}^{(2)}+\mathcal{O}(\theta^{4}), \label{eq:_DO_and_BF}
\end{flalign}
where, as shown in the Supplemental Material (SM)~\cite{SM1}, the second cumulant $\mathcal{C}_{M}^{(2)}$—i.e., the bipartite fluctuations—is given by
\begin{flalign}
\mathcal{C}_{M}^{(2)}=C_{J}\left[\frac{\pi}{12}\left(\frac{6l^{2}}{\epsilon^{2}}-\frac{12l}{\epsilon}\right)+\ln\left(\frac{l}{\epsilon}\right)+\mathcal{O}(1)\right].
\label{eq:_2nd_cumulant}
\end{flalign}
Here, $C_{J}$ is the current central charge, a universal characteristic of the CFT, and $\epsilon$ denotes a short-distance (real-space) UV cutoff. Equation~\eqref{eq:_2nd_cumulant} corresponds to the cubic special case of the more general trihedral-corner geometry, whose angular dependence for arbitrary trihedral corners is discussed in Ref.~\cite{wu2026corner}. The universal coefficient of the logarithmic term also appeared in Ref.~\cite{Dirac_log} in the context of Weyl semimetals. We emphasize that, for a cubic cut, the smooth part of $\partial M$ is completely flat, and therefore there is no curvature contribution to the logarithmic term, in contrast to shapes such as the sphere or cylinder.

To establish the disorder operator as a practical lattice probe of universal CFT data in three spatial dimensions, especially through its trihedral-corner logarithmic contribution, we focus on the conventional O(3) transition in this work. At the Gaussian fixed point, the current-current correlation takes the form in Eq.~\eqref{eq:_CFT_JJ} with $C_{J}=\frac{2}{d-1}[\Gamma(\frac{d+1}{2})/2\pi^{\frac{d+1}{2}}]^{2}$~\cite{petkou1996conserved,1994implications,diab2016on}, where $d$ is the spatial dimension. For the present three-dimensional quantum system, the corresponding field theory lies at the upper critical dimension,  where marginally irrelevant interactions are expected to introduce corrections to Gaussian fixed-point physics~\cite{Book_Zinn-Justin}. As we show in SM~\cite{SM1}, however, the logarithmically slow renormalization group (RG) flow of the interaction does not modify the coefficient of the logarithmic term in Eq.~\eqref{eq:_2nd_cumulant}. Consequently, in our numerical analysis of the corner term, we directly compare with the mean-field value $C_{J}=1/(4\pi^{4})$.


Another interesting limit arises from a small deformation of the region $M$ in a generic CFT.
Assuming that the disorder operator corresponds to a conformal surface defect in the IR,
a small deformation $\delta M$ yields $\langle X_{M+\delta M}\rangle=\langle X_{M}\rangle+\delta^{(2)}\langle X_{M}\rangle+\ldots$, where the leading correction $\delta^{(2)}\langle X_{M}\rangle$ is controlled by the displacement operator $D^{i}$ as follows~\cite{wang2021scaling,Billo2013line,Billo2016defects}
\begin{flalign}
\frac{1}{2}\int_{\partial M}\textrm{d}^{2}\boldsymbol{x}_{1}\int_{\partial M}\textrm{d}^{2}\boldsymbol{x}_{2}\langle D^{i}(\boldsymbol{x}_{1})D^{j}(\boldsymbol{x}_{2})\rangle\xi^{i}(\boldsymbol{x}_{1})\xi^{j}(\boldsymbol{x}_{2}),
\label{eq:_small_def}
\end{flalign}
where $\xi^{i}(\boldsymbol{x})$ parametrizes the deformation $\delta M$. The two-point function $\langle D^{i}(\boldsymbol{x})D^{j}(0)\rangle=C_{D}\delta^{ij}|\boldsymbol{x}|^{-6}$ is fixed by the Ward identity, and the universal coefficient $C_{D}$ is referred to as the defect central charge. Namely, as shown in SM~\cite{SM1}, in the symmetric flat limit of a trihedral corner—where the corner angles between any pair of edges are equal (denoted by $0<\phi<2\pi/3$)—the logarithmic term vanishes as $\delta^{(2)}\langle X_{M}\rangle\propto C_{D}(\phi-2\pi/3)\ln(l)$. A similar calculation in the context of entanglement geometry has been discussed in Ref.~\cite{4D_corner_5}.

In the following sections, we present the QMC results for $|\langle X(\theta)\rangle|$ for a cubic subregion in the (3+1)D Heisenberg model  (charge $n_i=S_i^z-\tfrac{1}{2}$ in the spin-1/2 Heisenberg model), and confirm that they are consistent with Eq.~\eqref{eq:_DO_fit}, successfully extracting $C_J$ from the relation $s=-\tfrac{C_J}{2}\theta^2$ as $\theta \to 0$. A detailed numerical investigation of the small deformation in Eq.~\eqref{eq:_small_def} for microscopic models is left for future work.

\begin{figure}
    \centering
    \includegraphics[width=1\linewidth]{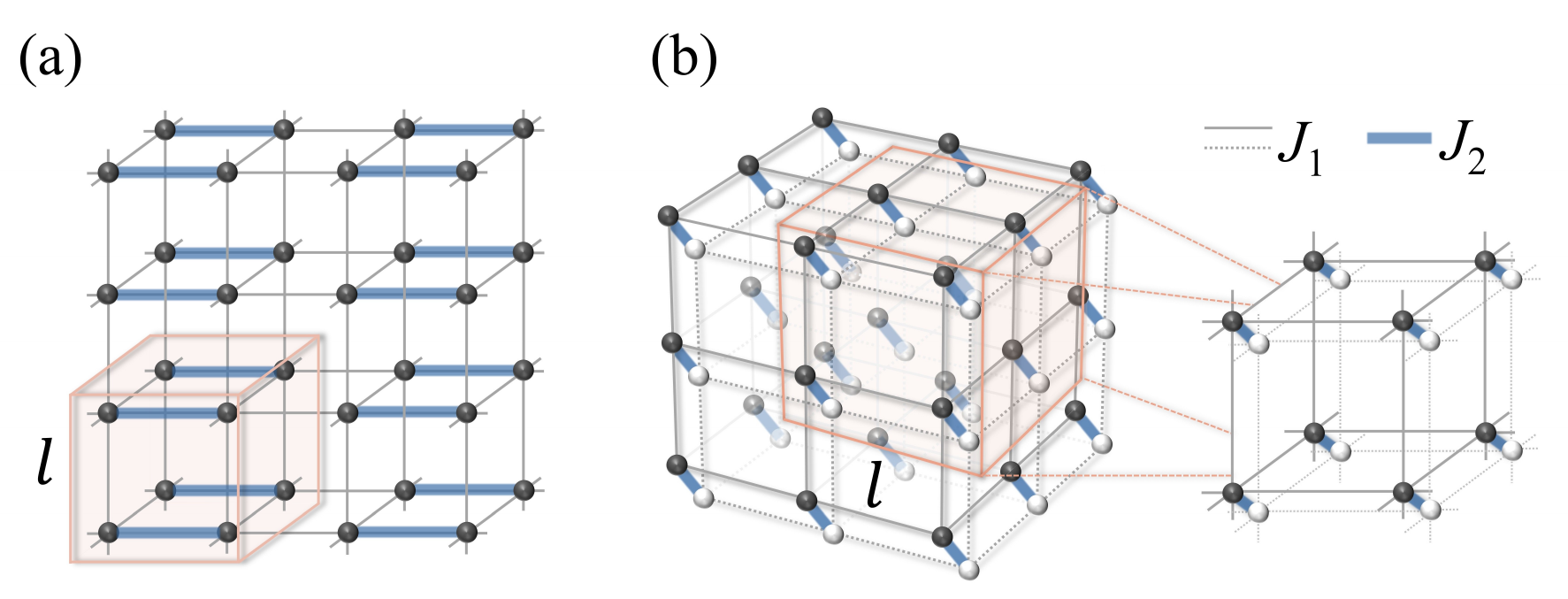}
    \caption{The two lattice models, (a) columnar-dimerized and (b) double-cubic AFM Heisenberg models with periodic boundary conditions. The weak interaction $J_1$ and the strong interaction $J_2$ are represented by thin and thick bonds. Orange cubes $M$ represent the region where the disorder operator is applied.} 
    \label{fig:Model}
\end{figure}

\begin{figure}
    \centering
    \includegraphics[width=0.7\linewidth]{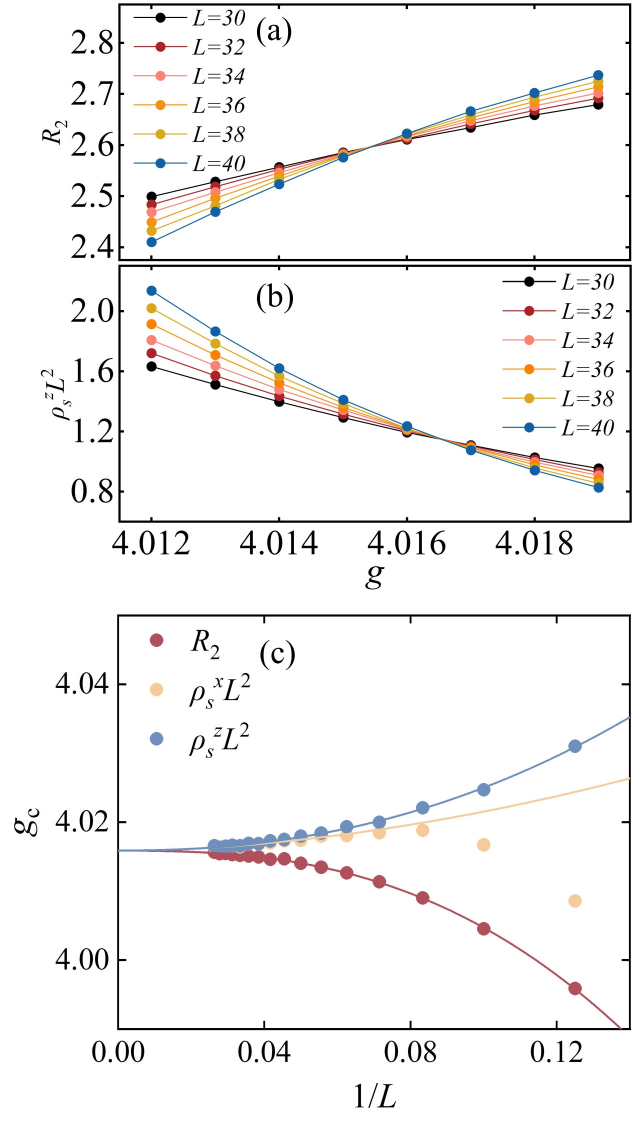}
    \caption{(a) The Binder ratio $R_2$ and (b) scaled spin stiffness $\rho_s^\alpha L^2$ in the $z$ direction for system sizes $L=30,32,...,40$ vs the coupling ratio g. (c) The size dependence of critical point g$_c^{R_2}(L)$ and g$_c^{\rho_s^\alpha}(L)$ obtained from the crossing of the curves of $R_2$ and $\rho_s^\alpha L^2(\alpha=x,z)$ with systems sizes $L$ and $L+2$. The curves represent the fitting function g$_c(L)$=g$_c$+$kL^{-(1/\nu+\omega)}$, where $\nu$=$\frac{1}{2}$ is the critical exponent of mean-field theory, and both fits indicate that g$_c$=4.0159(1) as $L\to\infty$.}
    \label{fig:criticalpoint}
\end{figure}

\mysec{Models and method.} We study the disorder operator of the following two models via stochastic series expansion (SSE) QMC simulations~\cite{sandvik1998stochastic,sandvik2010computational,Syljuaasen2002,yan2019sweeping,yan2020improved,sandvik2019stochastic}. The first is a CD antiferromagnetic (AFM) Heisenberg model with Hamiltonian
\begin{equation}
\begin{array}{l}
H_{CD}=J_1\sum\limits_{\left \langle i,j\right \rangle}{{{S}}_{i} \cdot {{S}}_{j}}+J_2\sum\limits_{\left \langle i,j \right \rangle'}{{S}}_{i} \cdot {{S}}_{j},
\end{array}
\label{Eq:Hmlt}
\end{equation}
where $S_i$ denotes the spin-1/2 operator on each site $i$; $\left \langle i,j\right \rangle$  and $\left \langle i,j \right \rangle'$represent the nearest-neighbor sites, as shown in FIG. \ref{fig:Model}(a). Previous work has revealed that quantum critical point (QCP) g$_c$($J_2/J_1$)=4.013(3)~\cite{ nohadani2005quantum}. To access reliable scaling behavior and universal quantities at QCP, we employ the size independent parameters Binder ratio $R_2=\left \langle m_{sz}^4 \right \rangle / \left \langle m_{sz}^2 \right \rangle^2$($m_{sz}^2$ is the staggered magnetization), as well as the scaled spin stiffness $\rho_s^\alpha L^2$ ($\alpha$ represents the direction of lattice) for all even-length system sizes $L= 8,10, ...,40$ to accurately identify the QCP, since the spin stiffness obeys the scaling law $\rho_s^\alpha\sim L^{2-d-z}$, where the $\rho_s^\alpha$ in the SSE simulation is defined as~\cite{sandvik2010computational}

\begin{equation}
\begin{array}{l}
\rho^\alpha_s=\frac{3}{2\beta L^3} \left\langle(N_{\alpha}^{+}-N_{\alpha}^{-})^{2} \right \rangle, 
\end{array}
\label{Eq:stiffness2}
\end{equation}
Here, $N_\alpha^+$ and $N_\alpha^-$ represent the number of off-diagonal transporting spin in the positive and negative direction, spatial dimension is $d=3$, and the dynamic exponent is $z=1$ in our case. In Figs.~\ref{fig:criticalpoint} (a) and (b), the crossing points of different sizes display that finite-size scaling is well converged for $R_2$ and $\rho_s^z L^2$ . According to the standard finite-size scaling theory, such size-dependent critical points $g_{c}(L)$ are expected to approach the true critical point $g_{c}$ as

\begin{equation}
\begin{array}{l}
 g_c(L)=g_c+kL^{-(1/\nu+\omega)}, 
\end{array}
\label{Eq:extrapolation}
\end{equation}
where $\omega$ is the effective correction exponent and $\nu$ is the correlation-length exponent. Notably, the logarithmic corrections at $D=D_c$ in our case suggest that Eq.~\eqref{Eq:extrapolation} should be modified to~\cite{Kenna2014Fisher,kenna2013new,qin2015multiplicative}

\begin{equation}
\begin{array}{l}
 g_c(L)=g_c+kL^{-(1/\nu+\omega)}ln^{\hat{c}}L, 
\end{array}
\label{Eq:extrapolation2}
\end{equation}
where $\hat{c}$ represents the exponent of logarithmic correction~\cite{qin2015multiplicative}. However, our simulations are limited to a finite range of system sizes, over which the dominant behavior can be well captured by the simpler fitting form of Eq.~\eqref{Eq:extrapolation}. After extracting all crossing points between system sizes $L$ and $L+2$, we show the finite-size extrapolation results of $R_2$ and $\rho_s^\alpha L^2$$(\alpha=x,z)$ to obtain a more precise QCP, where both fitting curves converge to the critical point g$_c$=4.0159(1) for $L\to\infty$ shown in Fig.~\ref{fig:criticalpoint} (c), which belongs to the (3+1)D O(3) universality class. 

The other model is the DC AFM Heisenberg model. The Hamiltonian is written as
\begin{equation}
\begin{array}{l}
H_{DC}=J_1\sum\limits_{\left \langle i,j\right \rangle}({{{S}}_{1,i} \cdot {{S}}_{1,j}}+{{{S}}_{2,i} \cdot {{S}}_{2,j}})+J_2\sum\limits_{i}{{S}}_{1,i} \cdot {{S}}_{2,i},
\end{array}
\label{Eq:Hmlt}
\end{equation}
where $J_2$ represents the inter-cube AFM interaction, as shown in Fig.~\ref{fig:Model}(b). The QCP between the AFM  N\'{e}el phase and the spin singlet phase is g$_c$=4.83704(6)~\cite{qin2015multiplicative}, which also belongs to the (3+1)D O(3) universality class. In principle, both models should contain identical CFT information extracted by the disorder operator at QCP.

\begin{figure}
    \centering
    \includegraphics[width=1.0\linewidth]{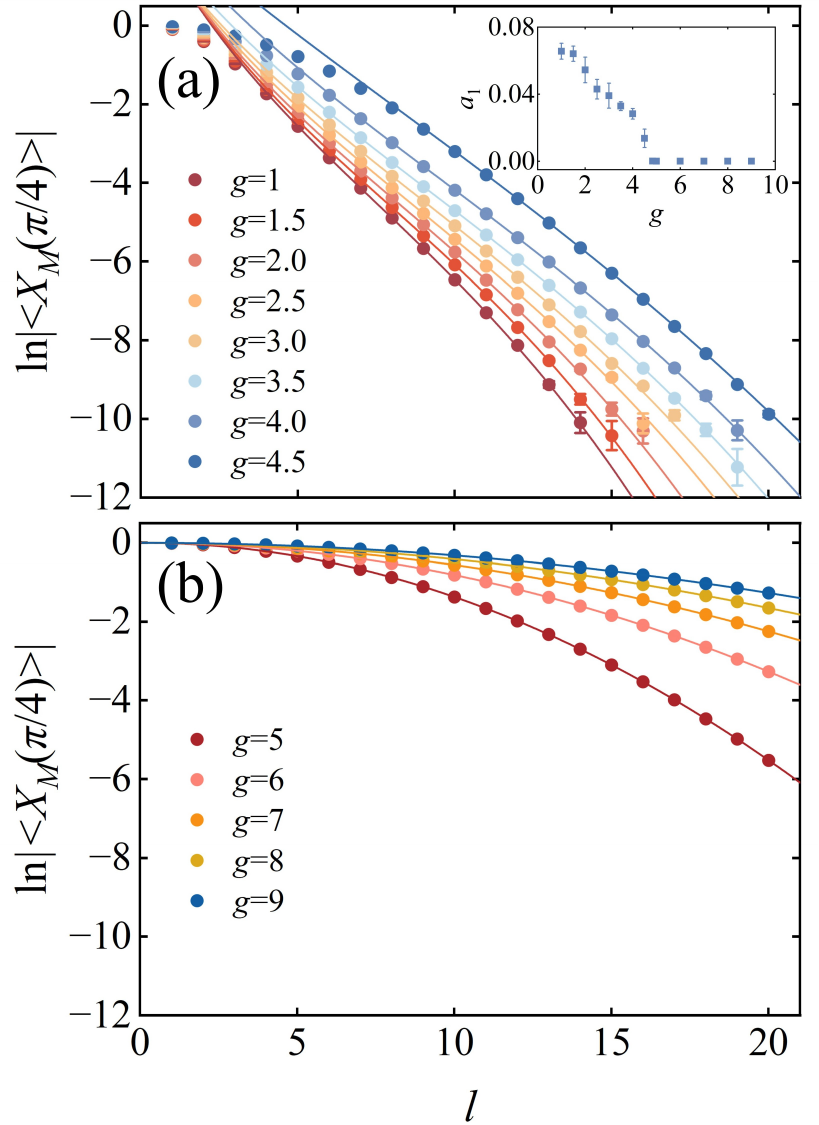}
    \caption{Disorder operator $\ln|\left \langle X_M(\theta=\pi/4)\right \rangle|$ in the (a) AFM N\'{e}el phase and (b) spin singlet phase for the double-cubic model with system size $L=40$ obtained by QMC, where $l$ is the size length of cube $M$. The solid lines represent the fitting function $\ln|\left \langle X_M(\theta)\right \rangle|=-a_{1}l^2\ln l+b_{1}l^2+c_{1}l+d$ for points $l\geq5$ to avoid the finite size effect in the AFM N\'{e}el phase, and $\ln|\left \langle X_M(\theta)\right \rangle|=-a_{2}l^2+b_{2}l+c_{2}$ in the spin singlet phase. The coefficient $a_1$ as a function of g is shown in the inset of (a).}
    \label{fig:anti}
\end{figure}

\mysec{Numerical results.} We adopt the inverse temperature $\beta=2L$ in the following QMC simulations and define the disorder operator $X_M(\theta)=\prod_M e^{i\theta (S_i^z-\tfrac{1}{2})}$ in the $l\times l \times l$ cube region within the lattice, as shown in Fig.~\ref{fig:Model}. For the CD model, it is preferable to avoid cutting strong bonds at the surfaces of the cube, as this increases the contribution of the leading term, leading to an amplification of the finite size error~\cite{wang2022scaling}. 

As shown in Fig.~\ref{fig:anti}, we obtain the expectation of the disorder operator with system size $L=40$. In the N\'{e}el phase, the spontaneous breaking of continuous symmetry leads to robust long-range AFM order, which is strongly manifested in the scaling behavior of the disorder operator, as presented in Fig.~\ref{fig:anti}(a). The expectation value of the disorder parameter decays extremely rapidly as $l$ increases, which is quantitatively well described by the fitting formula $\ln|\langle X(\theta) \rangle| = -a_1 l^2 \ln l + b_1 l^2 + c_1 l + d$. The leading term $-a_{1}l^2\ln l$ captures the AFM N\'{e}el order with continuous symmetry breaking, similar to the (2+1)D case. As shown in the inset of Fig.~\ref{fig:anti}(a), the coefficient $a_1$ reflects the contribution from long-range order, systematically vanishes as g approaches the QCP. This logarithmic enhancement of the scaling arises from stronger density correlations that decay more slowly in space, induced by the presence of Goldstone modes~\cite{BF_gas_1998,BF_gas_2007,BF_QCP_2012,Lake_2018,wang2022scaling,wang2021scaling}.


In the spin singlet phase, the expectation of the disorder operator obeys the normal relation $\ln|\left \langle X(\theta)\right \rangle|=-a_{2}l^2+b_{2}l+c_{2}$ as expected, as shown in Fig.~\ref{fig:anti}(b). The leading term $-a_2 l^2$ reflects the presence of a spin gap in the quantum disordered phase. Notably, as g increases further into the deep spin singlet phase, the decay of the disorder operator becomes slower, indicating that the ground state is increasingly insensitive to the imposed symmetry twist, due to the preserved SU(2) symmetry. 


\begin{figure}
    \centering
    \includegraphics[width=1\linewidth]{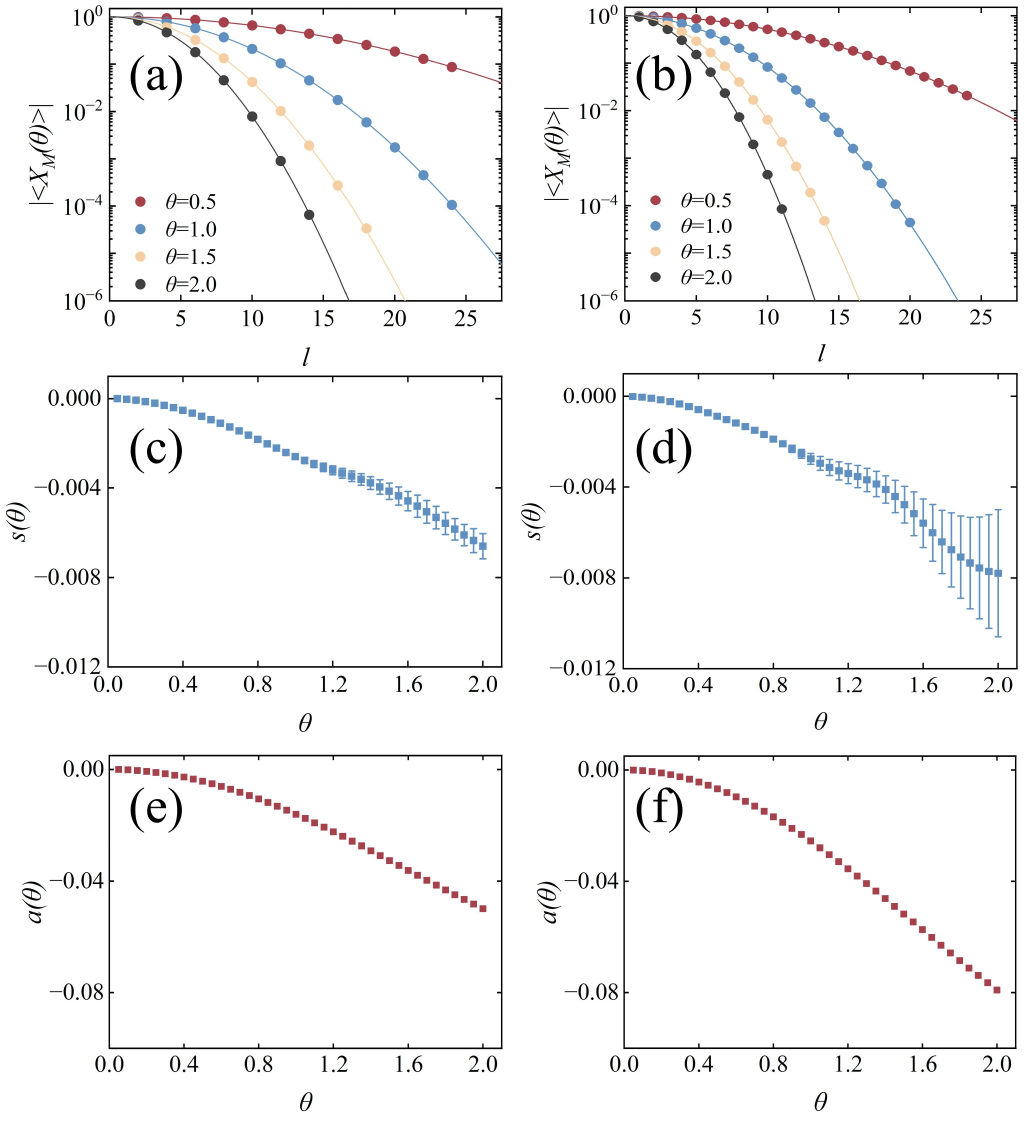}
    \caption{Disorder operator $|\left \langle X_M(\theta)\right \rangle|$ as a function of the cube with side length $l$ at QCP for $\theta$=0.5, 1.0, 1.5, 2.0 for (a) columnar-dimerized model and (b) double-cubic model with system size $L=48$ obtained by QMC. (c), (d) and (e), (f) show the logarithmic term coefficient $s(\theta)$ and leading quadratic term coefficient $a(\theta)$ extracted fromcolumnar-dimerized (CD) and double-cubic (DC) models, respectively.}
    \label{fig:QCP}
\end{figure}

We now investigate the scaling behavior of the disorder operator in the vicinity of the QCP in (3+1)D systems. To corroborate the universality of the logarithmic term coefficient predicted by CFT in Eq.~\eqref{eq:_DO_fit}, we compare the results obtained for both the CD and DC models. The expectation of the disorder operator with the system size $L=48$ at QCP are shown in Figs.~\ref{fig:QCP} (a) and (b). Remarkably, we find that the scaling form in Eq.~\ref{eq:_DO_fit} provides an excellent fit to the QMC data, even in the large $\theta$, capturing both the dominant quadratic decay of the disorder operator and the logarithmic correction arising from the trihedral-corner contribution. In Figs.~\ref{fig:QCP} (c) and (d), we show the logarithmic term coefficient $s(\theta)$ extracted from Eq.~\eqref{eq:_DO_fit}, and it is noteworthy that the statistical error in $s(\theta)$ is significantly larger in the DC model than in the CD model, especially at large $\theta$. This increased error primarily results from the more prominent contribution of the nonuniversal quadratic term in the DC model, which amplifies the statistical error in the fitted logarithmic term, since the dominant quadratic term can mask the subleading correction, as is further illustrated in Figs.~\ref{fig:QCP} (e) and (f), where the extracted coefficient $a(\theta)$ of the quadratic term is substantially larger in the DC model than in the CD model across the range of $\theta$. 
Despite these differences, both models display the same qualitative trends, further reinforcing the universality of the scaling structure at the (3+1)D QCP. 
\begin{figure}
    \centering
    \includegraphics[width=1\linewidth]{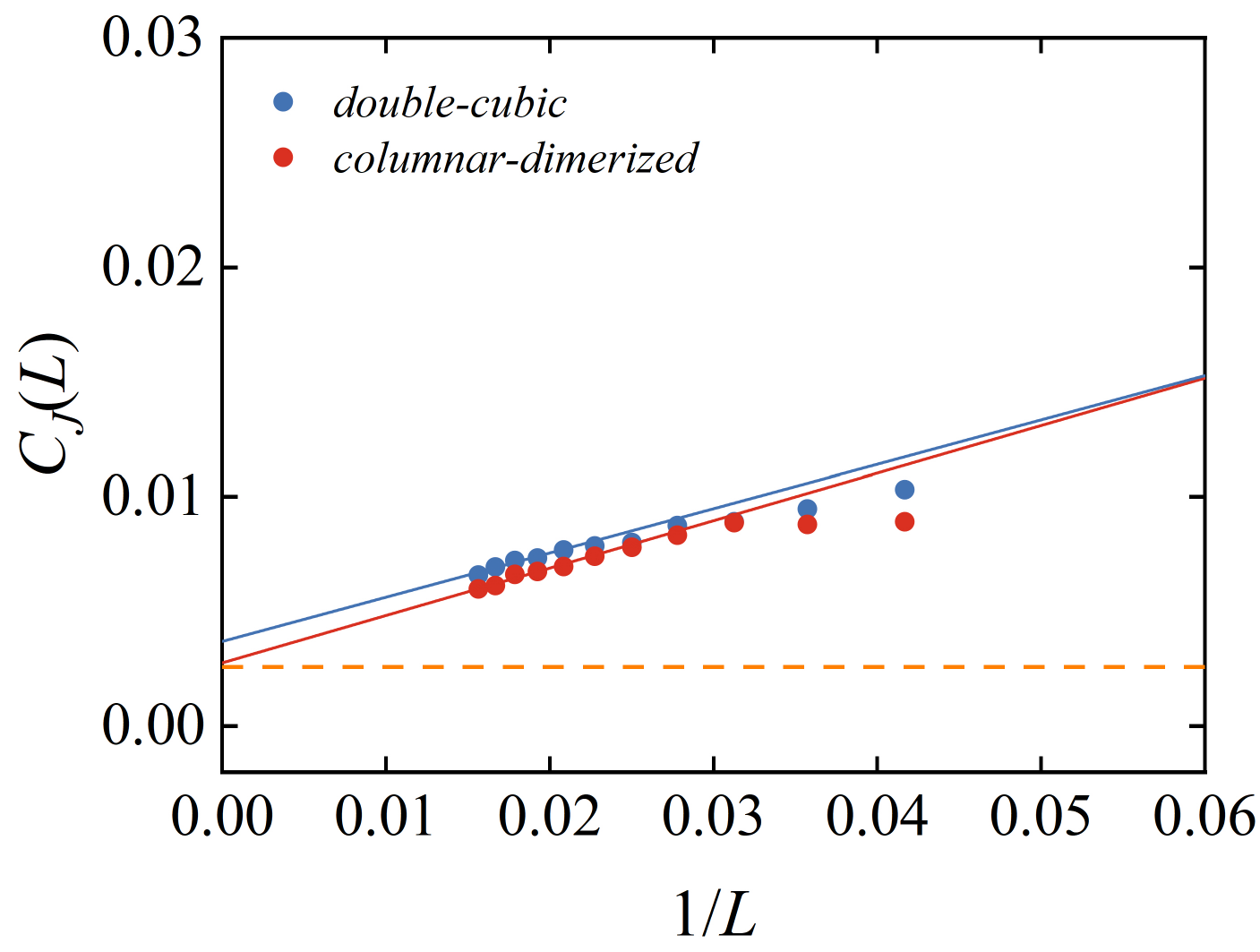}
    \caption{The finite-size extrapolation of the current central charge $C_J$ extracted from the disorder operator at QCP, with system sizes up to $L=64$ for the columnar-dimerized model and the double-cubic model. The orange dashed line represents the exact $C_J$ of the free theory in (3+1)D CFT.}
    \label{fig:CJ}
\end{figure}

In the case of small $\theta$($\le$ 0.3), the $C_J$ can be obtained by fitting the coefficient $s(\theta)$ by $s=-\tfrac{C_J}{2}\theta^2$ as expected. To assess the universal properties and compare with theoretical predictions, we perform finite-size extrapolation of $C_J$ for both the CD and DC models for system sizes up to $L=64$, as shown in Fig.~\ref{fig:CJ}. For the CD model, the extrapolated value of $C_J$ in the thermodynamic limit is $0.0028(4)$, in excellent agreement with the theoretical prediction of the free theory value $C_J = 0.00257$ as discussed above. This remarkable consistency between the theoretical prediction and the extrapolated value from the CD model provides strong evidence that our numerical method can accurately capture the universal behavior predicted by CFT at the (3+1)D QCP. In contrast, the extrapolated $C_J$ for the DC model is $0.0037(5)$, which is larger than the theoretical value. We attribute this difference primarily to stronger finite-size effects and the larger dominant quadratic term. As discussed above, the larger dominant quadratic term makes the subleading logarithmic contribution more difficult to isolate, and introduces systematic uncertainty in the extrapolation. Nevertheless, the DC result still supports the same universal trihedral-corner logarithmic contribution, while the less stable extrapolation reflects the stronger nonuniversal finite-size corrections in this model. Overall, our theoretical and computational analyses show that the disorder operator provides an efficient route to access universal CFT information in $(3+1)$D quantum systems.


\mysec{Summary and discussion.} To summarize, we have employed unbiased QMC simulations combined with theoretical calculations to study the scaling behavior of the disorder operator across the (3+1)D O(3) symmetry-breaking transition in two different microscopic spin models on the cubic lattice. We find good agreement with continuum field-theory predictions, including a quantitative determination of the current central charge at quantum criticality from the small-$\theta$ limit of the disorder operator associated with a cubic subregion. 
The verification of the universal formulas in the conventional O(3) transition fills in
the gap in numerical studies about the disorder operator in high dimensions and provides a useful protocol for future studies of unconventional quantum criticality. In particular, they may guide the search for candidate lattice models for the (3+1)D deconfined quantum critical points proposed in Ref.~\cite{Bi_4D_2019}. On another front, in the (2+1)D quantum systems, recent studies~\cite{Corner_QM_1,Corner_QM_2} have revealed an intriguing relation between the small-$\theta$ limit of the disorder operator and many-body quantum geometry. It would be interesting to explore the role of quantum geometry in the (3+1)D disorder operator in future work. For experiment, as the disorder operator is easily to measure in cold-atom systems~\cite{shao2024antiferromagnetic,browaeys2020many,bluvstein2021controlling}, our result provides an access to extract the information of field theory in experiment.


\mysec{Acknowledgments.}
X.L., X.W. and Z.L. contributed equally to this work. We thank Meng Cheng for illuminating conversations. X.L. and D.X.Y. were supported by NKRDPC-2022YFA1402802, NSFC-92165204, NSFC-12494591, Guangdong Provincial Key Laboratory of Magnetoelectric Physics and Devices (2022B1212010008), Research Center for Magnetoelectric Physics of Guangdong Province (2024B0303390001), and Guangdong Provincial Quantum Science Strategic Initiative (GDZX2401010). X.W. was supported by the Simons Collaboration on Ultra-Quantum Matter, which is a grant from the Simons Foundation (651442), and the Simons Investigator Grant (566116) awarded to S. Ryu. Z.L. and Z.Y. acknowledge the start-up funding of Westlake University, the China Postdoctoral Science Foundation under Grants No.2024M762935 and NSFC Special Fund for Theoretical Physics under Grants No.12447119. X.L., Z.L., Z.W., Z.Y., and D.X.Y. thank the high-performance computing center of Westlake University, the Beijing PARATERA Tech Co.,Ltd., and National Supercomputer Center in Guangzhou for providing HPC resources.
\bibliography{Ref}

\clearpage
\newpage
\onecolumngrid
\setcounter{page}{1}

\appendix
\begin{center}
\textbf{\large Supplemental Materials for ``Scaling of the disorder operator at (3+1)D O(3) quantum criticality''}\\
\end{center}

\begin{center}
{
Xuyang Liang,$^{1}$
Xiao-Chuan Wu,$^{2,3,*}$
Zenan Liu,$^{4,5,6}$
Zhe Wang,$^{5,6}$
Zheng Yan,$^{5,6,\dagger}$
and Dao-Xin Yao$^{1,\ddagger}$
}\par
\vspace{0.4em}

{\itshape
$^{1}$Guangdong Provincial Key Laboratory of Magnetoelectric Physics and Devices,\\
State Key Laboratory of Optoelectronic Materials and Technologies,\\
Center for Neutron Science and Technology, School of Physics,\\
Sun Yat-Sen University, Guangzhou, 510275, China\\[0.25em]
$^{2}$Department of Physics, Princeton University, Princeton, NJ 08544, USA\\
$^{3}$Leinweber Institute for Theoretical Physics, University of Chicago, Chicago, IL 60637, USA\\
$^{4}$Department of Physics, South China University of Technology, Guangzhou 510640, China\\
$^{5}$Department of Physics, School of Science and Research Center for\\
Industries of the Future, Westlake University, Hangzhou 310030, China\\
$^{6}$Institute of Natural Sciences, Westlake Institute for Advanced Study, Hangzhou 310024, China
}\par
\end{center}

\section{Derivations for the small-$\theta$ limit} \label{app:_BF}

In this appendix, we provide detailed derivations of the small-$\theta$ limit of the disorder operator in Eq.~\eqref{eq:_DO_and_BF}.  

The current correlation function in CFTs has a rigid structure,
\begin{align}
    \langle J_{\mu}(x)J_{\nu}(0)\rangle=\frac{C_{J}}{\left|x\right|^{2d}}\left(\delta^{\mu\nu}-\frac{2x^{\mu}x^{\nu}}{\left|x\right|^{2}}\right),
    \label{eq:_CFT_JJ}
\end{align}
where $d=3$ is the spatial dimension and $C_{J}$ is the current central charge. We introduce a dual two-form gauge field to represent the conserved $\mathrm{U}(1)$ current,
\begin{align}
    J_{\mu}=\frac{\mathtt{i}}{4\pi}\varepsilon_{\mu\nu\rho\sigma}\partial_{\nu}b_{\rho\sigma}.
\end{align}
We find that the gauge-field propagator takes the form\footnote{Here, we use the standard anti-symmetrization notation, where square brackets indicate normalization by the number of terms. For example, $T^{[\mu\nu]}=\frac{1}{2}(T^{\mu\nu}-T^{\nu\mu})$, $T^{[\mu\nu\rho\sigma]}=\frac{1}{4!}(T^{\mu\nu\rho\sigma}\pm\cdots)$, with the sum taken over all permutations of indices with appropriate sign.}
\begin{align}
    \langle b_{\mu\nu}(x)b_{\rho\sigma}(0)\rangle=C_{J}\frac{2\pi^{2}}{3}\frac{1}{|x|^{4}}\left((1+\zeta)\delta^{[\mu[\rho}\delta^{\sigma]\nu]}-4\zeta\frac{\delta^{[\mu[\rho}x^{\sigma]}x^{\nu]}}{|x|^{2}}\right)=C_{J}\frac{2\pi^{2}}{3}\left(\frac{\delta_{[\mu[\rho}\delta_{\sigma]\nu]}}{|x|^{4}}-\frac{\zeta}{2}\delta_{[\mu[\rho}\partial_{\sigma]}\partial_{\nu]}\frac{1}{|x|^{2}}\right),
\end{align}
where $\zeta$ is fixed by a specific gauge choice. The $\textrm{U}(1)$ disorder operator is represented by the Wilson surface
\begin{align}
    X_{M}(\theta)=\exp\left(\frac{\mathtt{i}\theta}{2\pi}\oint_{\partial M}\frac{1}{2}b_{\mu\nu}\textrm{d}x^{\mu}\wedge\textrm{d}x^{\nu}\right).
\end{align}
Both $\langle X_{M}(\theta)\rangle$ and its logarithm $\log\langle X_{M}(\theta)\rangle$ can be interpreted as generating functions. From them, the $n$-th moment and the $m$-th cumulant are extracted as
\begin{align}
    \mathcal{M}_{M}^{(n)}&=\lim_{\theta\rightarrow0}(-\mathtt{i}\partial_{\theta})^{n}\langle X_{M}(\theta)\rangle,\nonumber\\\mathcal{C}_{M}^{(m)}&=\lim_{\theta\rightarrow0}(-\mathtt{i}\partial_{\theta})^{m}\log\langle X_{M}(\theta)\rangle,
\end{align}
with $\mathcal{M}_{M}^{(n)}$ and $\mathcal{C}_{M}^{(m)}$ related through incomplete Bell polynomials. In CFTs, the one-point and three-point functions of any conserved current associated with an abelian symmetry vanish. As a result, the Taylor expansion of the moment-generating function $\langle\mathcal{W}_{\Sigma}(\vartheta)\rangle$ takes the form shown in Eq.~\eqref{eq:_DO_and_BF}.

The second cumulant $\mathcal{C}_{M}^{(2)}$ (i.e., the bipartite fluctuations) can be computed using the gauge field propagator
\begin{align}
    \mathcal{C}_{M}^{(2)}&=\frac{1}{(2\pi)^{2}}\oint_{\partial M}\textrm{d}S_{1}^{i}\frac{\varepsilon^{ijk}}{2}\oint_{\partial M}\textrm{d}S_{2}^{l}\frac{\varepsilon^{lmn}}{2}\langle b_{jk}(x_{1})b_{mn}(x_{2})\rangle.
\end{align}
To evaluate this expression, we need a gauge-invariant UV cutoff. For this purpose, we define the regularized function
\begin{align}
 &D_{il}^{\epsilon}(x,y,z)=\frac{1}{(2\pi)^{2}}\frac{\varepsilon^{ijk}}{2}\frac{\varepsilon^{lmn}}{2}\langle b_{jk}(\epsilon,x,y,z)b_{mn}(0,0,0,0)\rangle\nonumber\\&=\frac{C_{J}}{6}\left(\begin{array}{ccc}
\frac{(\zeta+1)(x^{2}+\epsilon^{2})+(1-\zeta)y^{2}+(1-\zeta)z^{2}}{2(x^{2}+y^{2}+z^{2}+\epsilon^{2})^{3}} & \frac{\zeta xy}{(x^{2}+y^{2}+z^{2}+\epsilon^{2})^{3}} & \frac{\zeta xz}{(x^{2}+y^{2}+z^{2}+\epsilon^{2})^{3}}\\
\frac{\zeta xy}{(x^{2}+y^{2}+z^{2}+\epsilon^{2})^{3}} & \frac{(1-\zeta)x^{2}+(\zeta+1)(y^{2}+\epsilon^{2})+(1-\zeta)z^{2}}{2(x^{2}+y^{2}+z^{2}+\epsilon^{2})^{3}} & \frac{\zeta yz}{(x^{2}+y^{2}+z^{2}+\epsilon^{2})^{3}}\\
\frac{\zeta xz}{(x^{2}+y^{2}+z^{2}+\epsilon^{2})^{3}} & \frac{\zeta yz}{(x^{2}+y^{2}+z^{2}+\epsilon^{2})^{3}} & \frac{(1-\zeta)x^{2}+(1-\zeta)y^{2}+(\zeta+1)(z^{2}+\epsilon^{2})}{2(x^{2}+y^{2}+z^{2}+\epsilon^{2})^{3}}
\end{array}\right),
\end{align}
where $\epsilon$ serves as a small real-space UV cut-off. 

We now take $\partial M$ to be the surface of a cube with side length $l$. The surface consists of six square faces, whose correlations fall into three classes. First, the self-correlation of a single face with normal $\hat{z}$ is
\begin{align}
    I_{1}&=\int_{0}^{l}\textrm{d}x_{1}\int_{0}^{l}\textrm{d}x_{2}\int_{0}^{l}\textrm{d}y_{1}\int_{0}^{l}\textrm{d}y_{2}\frac{(1-\zeta)(x_{1}-x_{2}){}^{2}+(1-\zeta)(y_{1}-y_{2}){}^{2}+(\zeta+1)\epsilon^{2}}{2((x_{1}-x_{2}){}^{2}+(y_{1}-y_{2}){}^{2}+\epsilon^{2}){}^{3}}\nonumber\\&=\frac{\pi l^{2}}{2\epsilon^{2}}+\frac{\pi(\zeta-2)l}{2\epsilon}-(\zeta-1)\log\left(\frac{l}{\epsilon}\right)+\frac{(\zeta-1)\log(4)-(4+\pi)\zeta+\pi+6}{4}+\mathscr{O}(\epsilon).
\end{align}
Second, the correlation between two adjacent faces that meet along the $\hat{z}$ axis contributes
\begin{align}
    I_{2}&=\int_{0}^{l}\textrm{d}y_{1}\int_{0}^{l}\textrm{d}z_{1}\int_{0}^{l}\textrm{d}x_{2}\int_{0}^{l}\textrm{d}z_{2}\frac{-\zeta x_{2}y_{1}}{(x_{2}^{2}+y_{1}^{2}+(z_{1}-z_{2}){}^{2}+\epsilon^{2}){}^{3}}\nonumber\\&=-\frac{\pi\zeta l}{8\epsilon}+\frac{\zeta}{4}\log\left(\frac{l}{\epsilon}\right)+\frac{\zeta}{8}\left(2+\pi-\log\left(\frac{8}{3}\right)-\sqrt{2}\cot^{-1}(\sqrt{2})\right)+\mathscr{O}(\epsilon).
\end{align}
Third, the correlation between two opposite faces separated by a distance $l$ in the $\hat{z}$ direction is
\begin{align}
    I_{3}&=\int_{0}^{l}\textrm{d}x_{1}\int_{0}^{l}\textrm{d}x_{2}\int_{0}^{l}\textrm{d}y_{1}\int_{0}^{l}\textrm{d}y_{2}\frac{(\zeta-1)(x_{1}-x_{2}){}^{2}+(\zeta-1)(y_{1}-y_{2}){}^{2}-(\zeta+1)(l^{2}+\epsilon^{2})}{2(l^{2}+(x_{1}-x_{2}){}^{2}+(y_{1}-y_{2}){}^{2}+\epsilon^{2}){}^{3}}\nonumber\\&=-\frac{1}{4}\pi(\zeta-2)+\frac{1}{2}(\zeta-1)\log\left(\frac{4}{3}\right)+\frac{(\zeta-4)}{\sqrt{2}}\cot^{-1}(\sqrt{2}).
\end{align}
Because the cube has six faces, twenty-four pairs of adjacent faces, and six pairs of opposite faces, the total becomes
\begin{align}
    \frac{\mathcal{C}_{M}^{(2)}}{C_{J}}&=\frac{1}{6}(6I_{1}+24I_{2}+6I_{3})=\frac{\pi l^{2}}{2\epsilon^{2}}-\frac{\pi l}{\epsilon}+\log\left(\frac{l}{\epsilon}\right)+\frac{1}{4}\left(6+3\pi-2\log\left(\frac{8}{3}\right)-8\sqrt{2}\cot^{-1}(\sqrt{2})\right)+\mathscr{O}(\epsilon)\nonumber\\&=\frac{\pi}{12}\left(\frac{6l^{2}}{\epsilon^{2}}-\frac{12l}{\epsilon}\right)+\log\left(\frac{l}{\epsilon}\right)+\mathscr{O}(1),
\end{align}
where $6l^{2}$ is the total surface area of the cube, and $12l$ is the total edge length. The universal logarithmic term arises from the trihedral corners.

For completeness, let us also consider a few simple examples. We begin with the simplest case in which $\partial M$ is a perfect sphere of radius R in real space. For a general gauge choice $\zeta$, the second cumulant is
\begin{align}
    \mathcal{C}_{M}^{(2)}&=\int_{0}^{2\pi}\textrm{d}\phi_{1}\int_{0}^{\pi}\textrm{d}\theta_{1}R^{2}\sin\theta_{1}\int_{0}^{2\pi}\textrm{d}\phi_{2}\int_{0}^{\pi}\textrm{d}\theta_{2}R^{2}\sin\theta_{2}\left(\hat{\boldsymbol{n}}_{1}\cdot D^{\epsilon}(\boldsymbol{x}_{1}-\boldsymbol{x}_{2})\cdot\hat{\boldsymbol{n}}_{2}\right),\nonumber\\&\textrm{where}\qquad\begin{cases}
\boldsymbol{x}_{1}=R\hat{\boldsymbol{n}}_{1}=R(\sin\theta_{1}\cos\phi_{1},\sin\theta_{1}\sin\phi_{1},\cos\theta_{1})\\
\boldsymbol{x}_{2}=R\hat{\boldsymbol{n}}_{2}=R(\sin\theta_{2}\cos\phi_{2},\sin\theta_{2}\sin\phi_{2},\cos\theta_{2})
\end{cases}\!\!\!\!\!\!\!.
\end{align}
Thanks to rotational symmetry, we may first integrate over $\boldsymbol{x}_{1}$ while keeping $\boldsymbol{x}_{2}=(0,0,R)$ fixed. The remaining integral over $\boldsymbol{x}_{2}$ then contributes an overall factor of $4\pi R^{2}$. The final result is 
\begin{align}
    \frac{\mathcal{C}_{M}^{(2)}}{C_{J}}=\frac{\pi^{2}}{3}\left(\frac{R^{2}}{\epsilon^{2}}-\log\left(\frac{2R}{\epsilon}\right)+\frac{1}{4}\right)+\mathscr{O}(\epsilon)=\frac{\pi}{12}\frac{4\pi R^{2}}{\epsilon^{2}}-\frac{\pi^{2}}{3}\log\left(\frac{R}{\epsilon}\right)+\mathscr{O}(1),
\end{align}
which is independent of the gauge parameter $\zeta$. Here, $4\pi R^{2}$ is the total surface area.

We now turn to a cylinder of radius $R$ and length $L$. Its surface
consists of two circular end caps and a rectangular side wall, which
give rise to four distinct integrals. To simplify the calculation
we set the gauge parameter to $\zeta=0$. The end-cap self-correlation
gives 
\begin{flalign}
I_{1} & =\int_{0}^{2\pi}\textrm{d}\theta_{1}\int_{0}^{R}\textrm{d}r_{1}r_{1}\int_{0}^{2\pi}\textrm{d}\theta_{2}\int_{0}^{R}\textrm{d}r_{2}r_{2}\frac{1}{2(-2r_{1}r_{2}\cos(\theta_{1}-\theta_{2})+r_{1}^{2}+r_{2}^{2}+\epsilon^{2}){}^{2}}=\frac{\pi^{2}R^{2}}{2\epsilon^{2}}-\frac{\pi^{2}R}{2\epsilon}+\mathscr{O}(1).
\end{flalign}
There is also a UV-finite contribution from the correlation between
the two end caps
\begin{flalign}
I_{2} & =\int_{0}^{2\pi}\textrm{d}\theta_{1}\int_{0}^{R}\textrm{d}r_{1}r_{1}\int_{0}^{2\pi}\textrm{d}\theta_{2}\int_{0}^{R}\textrm{d}r_{2}r_{2}\frac{-1}{2(L^{2}-2r_{1}r_{2}\cos(\theta_{1}-\theta_{2})+r_{1}^{2}+r_{2}^{2}+\epsilon^{2}){}^{2}}\nonumber \\
 & =\pi^{2}\left(-\frac{R^{2}}{2L^{2}}+\frac{\sqrt{L^{2}+4R^{2}}}{4L}-\frac{1}{4}\right)+\mathscr{O}(\epsilon)=\mathscr{O}(1).
\end{flalign}
The side-wall self-correlation leads to 
\begin{flalign}
I_{3} & =\int_{0}^{L}\textrm{d}z_{1}\int_{0}^{2\pi}\textrm{d}\theta_{1}R\int_{0}^{L}\textrm{d}z_{2}\int_{0}^{2\pi}\textrm{d}\theta_{2}R\frac{\cos(\theta_{1}-\theta_{2})}{2(-2R^{2}\cos(\theta_{1}-\theta_{2})+2R^{2}+(z_{1}-z_{2}){}^{2}+\epsilon^{2}){}^{2}}\nonumber \\
 & =\frac{\pi^{2}LR}{\epsilon^{2}}-\frac{\pi^{2}R}{\epsilon}-\frac{3\pi^{2}L}{8R}\log\left(\frac{L}{\epsilon}\right)+\mathscr{O}(1).
\end{flalign}
Under the gauge choice $\zeta=0$, the end-cap-to-side-wall correlation
vanishes because the two surfaces meet at right angles. We denote
it by $I_{4}=0$. Combining the above pieces, we have 
\begin{flalign}
\frac{\mathcal{C}_{M}^{(2)}}{C_{J}} & =\frac{1}{6}(2I_{1}+2I_{2}+I_{3}+4I_{4})=\frac{\pi}{12}\left(\frac{2\pi R(L+R)}{\epsilon^{2}}-\frac{4\pi R}{\epsilon}\right)-\frac{\pi^{2}L}{16R}\log\left(\frac{L}{\epsilon}\right)+\mathscr{O}(1),
\end{flalign}
where $2\pi R(L+R)$ is the total surface area, and $4\pi R$ is the
total edge length.

The universal coefficients are consistent with the results of Ref.~\cite{Dirac_log}, obtained by a slightly different method in the context of Weyl semimetals. We emphasize that the logarithmic term in the cubic case originates from trihedral corners, while in the sphere and cylinder it arises from the nonzero curvature of the bipartition surface, in agreement with the Solodukhin formula~\cite{Solodukhin_2008}.

\section{Interaction Effects} 
\label{app:_interaction}

The $\mathrm{O}(N)$ symmetry-breaking transition in $(d+1)$ dimensions is described by the Euclidean action
\begin{flalign}
\mathcal{S}=\int\textrm{d}^{d+1}x\left(\frac{1}{2}|\partial_{\mu}\boldsymbol{\phi}|^{2}+\frac{r}{2}|\boldsymbol{\phi}|^{2}+\frac{u}{4!}|\boldsymbol{\phi}|^{4}+\ldots\right)
\end{flalign}
where $\boldsymbol{\phi}$ denotes the $N$-component order parameter. At the critical point $r=0$, treating the interaction $u$ perturbatively, one finds the current-current correlation function $\Pi_{\mu\nu}=\langle J^{\mu}J^{\nu}\rangle$~\cite{petkou1996conserved,1994implications,diab2016on}
\begin{flalign}
\Pi_{\mu\nu}(x)=\frac{C_{J,0}}{|x|^{6}}\left(1-\frac{N+2}{12}\tilde{u}^{2}+\mathscr{O}(\tilde{u}^{3})\right)\left(\delta^{\mu\nu}-\frac{2x^{\mu}x^{\nu}}{|x|^{2}}\right).
\label{eq:_JJ_int}
\end{flalign}
Here, $C_{J,0}$ is the current central charge at the Gaussian fixed point, and
$\tilde{u}$ denotes the rescaled coupling, 
\begin{flalign}
C_{J,0}=\frac{2}{d-1}\frac{1}{\Omega_{d}^{2}},\qquad\tilde{u}=\frac{1}{2}\frac{\Omega_{d}}{(2\pi)^{d+1}}u,
\end{flalign}
with $\Omega_{d}=\Gamma(\frac{d+1}{2})/2\pi^{\frac{d+1}{2}}$ the surface area of the unit $d$-sphere.

Under RG flow, the renormalized correlator
$\Pi_{\mu\nu}^{\textrm{R}}$ satisfies the Callan-Symanzik equation
\begin{flalign}
\left(\frac{\partial}{\partial\ell}+\beta(\tilde{u})\frac{\partial}{\partial\tilde{u}}+2\eta_{J}\right)\Pi_{\mu\nu}^{\textrm{R}}(x;\tilde{u},\ell)=0
\label{eq:_Callan-Symanzik_eq}
\end{flalign}
where $\eta_{J}=0$ is the vanishing anomalous dimension of the conserved current $J$.
The beta function for $\tilde{u}$ is~\cite{Book_Zinn-Justin}
\begin{align}
\beta(\tilde{u})\equiv\frac{\textrm{d}\tilde{u}}{\textrm{d}\ell}=(3-d)\tilde{u}-\frac{N+8}{3}\tilde{u}^{2}+\mathscr{O}(\tilde{u}^{3}),
\end{align}
where $\ell$ is the RG scale parameter.

For $d=3-\epsilon$, the RG flow equation admits a perturbative fixed point
\begin{align}
\tilde{u}_{*}=\frac{3\epsilon}{N+8}+\mathscr{O}(\epsilon^{2}),
\label{eq:_WF_fp}
\end{align}
which corresponds to the $\textrm{O}(N)$ Wilson-Fisher fixed point. Substituting the fixed-point value given in Eq.~\eqref{eq:_WF_fp} into Eq.~\eqref{eq:_JJ_int}, we obtain the current central charge $C_{J}$ with interaction corrections~\cite{petkou1996conserved,1994implications,diab2016on}
\begin{align}
&\Pi_{\mu\nu}(x)=\frac{C_{J}}{|x|^{6}}\left(\delta^{\mu\nu}-\frac{2x^{\mu}x^{\nu}}{|x|^{2}}\right),\nonumber\\&C_{J}=C_{J,0}\left(1-\frac{3(N+2)}{4(N+8)^{2}}\epsilon^{2}+\mathscr{O}(\epsilon^{3})\right).
\end{align}

At the upper critical dimension $d=3$, the coupling $\tilde{u}$ becomes marginally irrelevant and flows logarithmically slowly. It is convenient to introduce the running coupling function
\begin{align}
U(\tilde{u},\ell)=\frac{\tilde{u}}{1+\frac{N+8}{3}\tilde{u}\ell}
\end{align}
which describes the solution of the beta function and satisfies 
\begin{align}
\frac{\partial U(\tilde{u},\ell)}{\partial\ell}=\beta(U(\tilde{u},\ell))=\beta(\tilde{u})\frac{\partial U(\tilde{u},\ell)}{\partial\tilde{u}},\qquad U(\tilde{u},0)=\tilde{u}.
\end{align}
The solution of the Callan-Symanzik equation Eq.~\eqref{eq:_Callan-Symanzik_eq} can then be written in the standard form
\begin{align}
\Pi_{\mu\nu}^{\textrm{R}}(x;\tilde{u},\ell)=[Z_{J}^{-1}(\tilde{u},-\ell)]^{2}\Pi_{\mu\nu}(x;U(\tilde{u},-\ell),0),
\end{align}
where the renormalization factor $Z_{J}$ is trivial due to $\eta_{J}=0$
\begin{align}
Z_{J}=\exp\left(-\int_{0}^{\ell}\textrm{d}s\eta_{J}\right)=1.
\end{align}
Matching the RG scale as $\ell=\log(\mu|x|)$, the large-$\ell$  behavior becomes
\begin{align}
&\Pi_{\mu\nu}^{\textrm{R}}(x)=\frac{C_{J}(x)}{|x|^{6}}\left(\delta^{\mu\nu}-\frac{2x^{\mu}x^{\nu}}{|x|^{2}}\right)\nonumber\\&C_{J}(x)=C_{J,0}\left(1-\frac{3(N+2)}{4(N+8)^{2}}\left(\frac{1}{\log(\mu|x|)}\right)^{2}+\ldots\right).
\end{align}
The (equal-time) static structure factor is given by $S(\boldsymbol{r})=-\Pi_{\tau\tau}^{\textrm{R}}(\tau\rightarrow0,\boldsymbol{r})$. The contribution that controls the corner-induced logarithmic term is then (see our follow-up work [] for explanations)
\begin{align}
\int_{\delta}^{l}\textrm{d}r(-r^{5})S(r)=C_{J,0}\left(\log(l/\delta)-\frac{3(N+2)}{4(N+8)^{2}}\frac{\log(l/\delta)}{\log(\mu l)\log(\mu\delta)}+\ldots\right).
\end{align}
We therefore conclude that marginal interactions at the upper critical dimension leave the coefficient of the logarithmic contribution in Eq.~\eqref{eq:_2nd_cumulant} unchanged, retaining its Gaussian fixed-point value.

\section{Planar limit of a  trihedral corner} \label{app:_trihedral}

\begin{figure}
    \centering
    \includegraphics[width=0.45\linewidth]{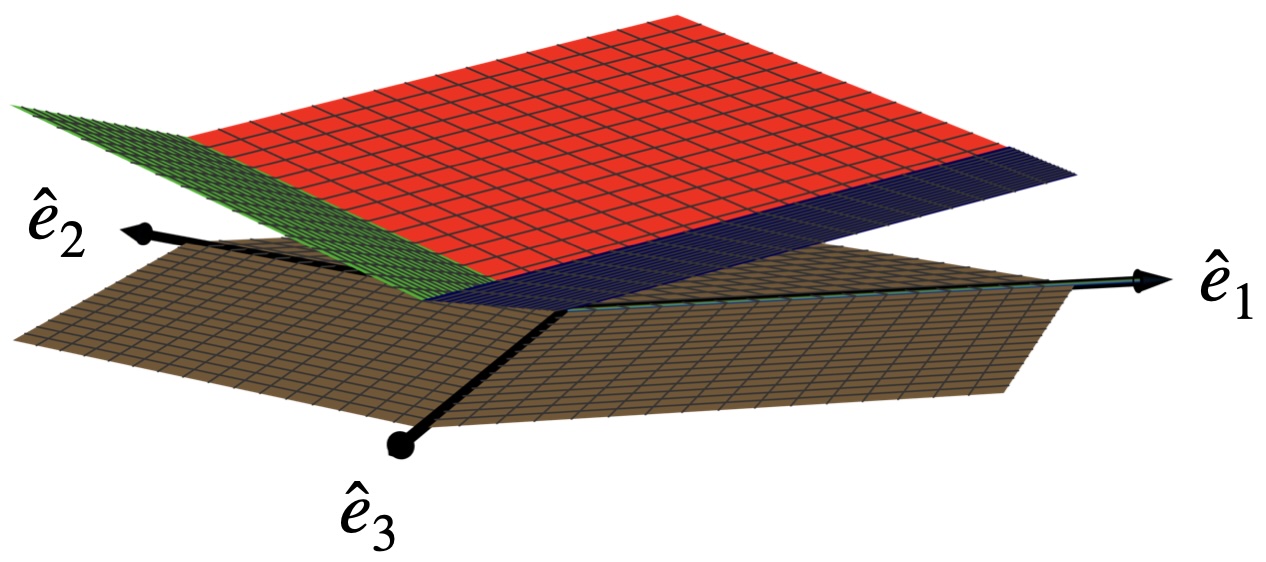}
    \caption{Trihedral corner deformation of a planar surface defect.}
    \label{fig:_trihedral}
\end{figure}

In this appendix, we consider a trihedral corner deformation of a planar surface defect, as schematically illustrated in  Fig.~\ref{fig:_trihedral}, which is generally characterized by the three angles $\phi_{1},\phi_{2},\phi_{3}$ between the three pairs of edges. For simplicity, we focus on the symmetric flat limit where $\phi_{1}=\phi_{2}=\phi_{3}\equiv\phi$ and $\phi$ is close to $2\pi/3$. The total contribution in Eq.~\eqref{eq:_small_def} consists of two types of integrals, 
\begin{flalign}
\delta^{(2)}\langle X_{M}\rangle=\frac{C_{D}}{2}(3I_{\textrm{same}}+6I_{\textrm{diff}}).
\end{flalign}

To parametrize the three planes, we introduce the following unit vectors:
\begin{flalign}
\hat{\boldsymbol{e}}_{1}=\left(\begin{array}{c}
1\\
0\\
0
\end{array}\right),\qquad\hat{\boldsymbol{e}}_{2}=\left(\begin{array}{c}
-1/2\\
\sqrt{3}/2\\
0
\end{array}\right),\qquad\hat{\boldsymbol{e}}_{3}=\left(\begin{array}{c}
-1/2\\
-\sqrt{3}/2\\
0
\end{array}\right).
\end{flalign}
The deformed plane can then be expressed as 
\begin{equation}
\boldsymbol{r}=x\hat{\boldsymbol{e}}_{i}+y\hat{\boldsymbol{e}}_{j}+\xi(x,y)\hat{z}
\end{equation}
where $x,y>0$ and $\hat{z}=(0,0,1)^{\mathsf{T}}$. Each pair of unit
vectors $\hat{\boldsymbol{e}}_{i},\hat{\boldsymbol{e}}_{j}$ specifies
one of the three intersecting surfaces forming the corner. The small
deformation is parametrized as
\begin{equation}
\xi(x,y)=\left.(x\hat{\boldsymbol{e}}_{i}+y\hat{\boldsymbol{e}}_{j})\cdot\frac{\hat{\boldsymbol{e}}_{i}+\hat{\boldsymbol{e}}_{j}}{|\hat{\boldsymbol{e}}_{i}+\hat{\boldsymbol{e}}_{j}|}\tan\alpha\right|_{i\neq j}=\frac{x+y}{2}\tan\alpha,
\end{equation}
where the angle variable $\alpha$ is related to $\phi$ through $\tan(\phi/2)=\sqrt{3}\cos\alpha$,
which gives $\alpha^{2}\approx4(2\pi/3-\phi)/\sqrt{3}$ in the small-angle
limit.

The contribution from the same surface can be computed as
\begin{flalign}
I_{\textrm{same}} & =|\hat{\boldsymbol{e}}_{1}\times\hat{\boldsymbol{e}}_{2}|^{2}\int_{0}^{l}\textrm{d}x_{1}\int_{0}^{l}\textrm{d}y_{1}\int_{0}^{l}\textrm{d}x_{2}\int_{0}^{l}\textrm{d}y_{2}\frac{\xi(x_{1},y_{1})\xi(x_{2},y_{2})}{|(x_{1}-x_{2})\hat{\boldsymbol{e}}_{1}+(y_{1}-y_{2})\hat{\boldsymbol{e}}_{2}|^{6}},
\end{flalign}
Introducing new variables $a_{1}=x_{1}+x_{2}$, $b_{1}=x_{1}-x_{2}$,
$a_{2}=y_{1}+y_{2}$, and $b_{2}=y_{1}-y_{2}$, we obtain 
\begin{flalign}
I_{\textrm{same}} & =\frac{3(\tan\alpha)^{2}}{16}\int_{-l}^{l}\textrm{d}b_{1}\int_{|b_{1}|}^{2l-|b_{1}|}\frac{\textrm{d}a_{1}}{2}\int_{-l}^{l}\textrm{d}b_{2}\int_{|b_{2}|}^{2l-|b_{2}|}\frac{\textrm{d}a_{2}}{2}\frac{(a_{1}+a_{2}+b_{1}+b_{2})(a_{1}+a_{2}-b_{1}-b_{2})}{4|b_{1}\hat{\boldsymbol{e}}_{1}+b_{2}\hat{\boldsymbol{e}}_{2}|^{6}}.
\end{flalign}
Since our focus is on the logarithmic term, we drop all power-law
contributions in $l$ and find
\begin{flalign}
I_{\textrm{same}} & \supset\frac{(\tan\alpha)^{2}}{64}\sum_{s_{1},s_{2}=\pm}\int_{0}^{l}\textrm{d}b_{1}\int_{0}^{l}\textrm{d}b_{2}\frac{b_{1}b_{2}(b_{1}^{2}+b_{2}^{2}-3(s_{1}b_{1}+s_{2}b_{2}){}^{2})}{|s_{1}b_{1}\hat{\boldsymbol{e}}_{1}+s_{2}b_{2}\hat{\boldsymbol{e}}_{2}|^{6}}.
\end{flalign}
Changing variables from $b_{1},b_{2}$ to a radial variable $\lambda$ and barycentric
coordinates $v_{1},v_{2}$ with $u_{i}=\lambda v_{i}$ and $v_{1}+v_{2}=1$, the logarithmic divergence arises from the $\lambda$-integral, yielding
\begin{equation}
I_{\textrm{same}}\supset-\left(\frac{1}{4}+\frac{5\pi}{72\sqrt{3}}\right)(\tan\alpha)^{2}\log(l).
\end{equation}

The contribution from a pair of distinct surfaces is
\begin{flalign}
I_{\textrm{diff}} & =(\hat{\boldsymbol{e}}_{1}\times\hat{\boldsymbol{e}}_{2})\cdot(\hat{\boldsymbol{e}}_{2}\times\hat{\boldsymbol{e}}_{3})\int_{0}^{l}\textrm{d}x_{1}\int_{0}^{l}\textrm{d}y_{1}\int_{0}^{l}\textrm{d}x_{2}\int_{0}^{l}\textrm{d}y_{2}\frac{\xi(x_{1},y_{1})\xi(x_{2},y_{2})}{|x_{1}\hat{\boldsymbol{e}}_{1}+(y_{1}-x_{2})\hat{\boldsymbol{e}}_{2}-y_{2}\hat{\boldsymbol{e}}_{3}|^{6}}.
\end{flalign}
Defining new variables $a=y_{1}+x_{2}$ and $b=y_{1}-x_{2}$, we find
\begin{flalign}
I_{\textrm{diff}} & =\frac{3(\tan\alpha)^{2}}{64}\int_{0}^{l}\textrm{d}x_{1}\int_{0}^{l}\textrm{d}y_{2}\int_{-l}^{l}\textrm{d}b\int_{|b|}^{2l-|b|}\frac{\textrm{d}a}{2}\frac{(2x_{1}+a+b)(2y_{2}+a-b)}{|x_{1}\hat{\boldsymbol{e}}_{1}+(y_{1}-x_{2})\hat{\boldsymbol{e}}_{2}-y_{2}\hat{\boldsymbol{e}}_{3}|^{6}}.
\end{flalign}
Dropping again all power-law contributions in $l$, we obtain
\begin{flalign*}
I_{\textrm{same}} & \supset\frac{(\tan\alpha)^{2}}{32}\int_{0}^{l}\textrm{d}x_{1}\int_{0}^{l}\textrm{d}y_{2}\int_{0}^{l}\textrm{d}b\sum_{s=\pm}\frac{b^{3}+3sb^{2}(x_{1}-y_{2})-6bx_{1}y_{2})}{|x_{1}\hat{\boldsymbol{e}}_{1}+sb\hat{\boldsymbol{e}}_{2}-y_{2}\hat{\boldsymbol{e}}_{3}|^{6}}.
\end{flalign*}
Using the radial-simplex decomposition with $x_{1}=\lambda v_{1}$, $y_{2}=\lambda v_{2}$, $b=\lambda v_{3}$ and $v_{1}+v_{2}+v_{3}=1$, the logarithmic divergence again originates from the $\lambda$-integral.
Performing the integration yields
\begin{equation}
I_{\textrm{diff}}\supset\frac{27+4\sqrt{3}\pi}{1728}(\tan\alpha)^{2}\log(l).
\end{equation}

Combining both contributions, we obtain
\begin{flalign}
\delta^{(2)}\langle X_{M}\rangle&\supset-\frac{C_{D}}{2}\left(\frac{21}{32}+\frac{\pi}{6\sqrt{3}}\right)(\tan\alpha)^{2}\log(l)\\&=-C_{D}\left(\frac{21}{16\sqrt{3}}+\frac{\pi}{9}\right)\left(\frac{2\pi}{3}-\phi\right)\log(l).
\end{flalign}
Comparing with the analogous calculation for entanglement entropy in Ref.~\cite{4D_corner_5}, we find that both results show the same scaling behavior, where the contribution of the deformed planar defect vanishes as $(2\pi/3-\phi)\log(l)$,
albeit with a slightly different overall coefficient. This discrepancy may originate from subtle differences between the infinite-square and infinite-hexagon geometries used to define the planar defect.

\end{document}